\documentclass[10pt,a4paper,twoside]{article}
\usepackage{epsfig}
\usepackage{baltlat6}
\usepackage{array}
\usepackage{here}
\begin{document}
\ \ \vspace{0.5mm} \setcounter{page}{1} \vspace{8mm}

\titlehead{Baltic Astronomy, vol.\,17, 1--8, 2011}

\titleb{THE INFLUENCE OF CHEMI-IONIZATION AND\\ 
RECOMBINATION PROCESSES ON SPECTRAL LINE SHAPES IN STELLAR
ATMOSPHERES}

\begin{authorl}
\authorb{Anatolij A. Mihajlov}{1,2}
\authorb{Ljubinko M. Ignjatovi{\' c}}{1,2}
\authorb{Vladimir A. Sre{\' c}kovi{\' c}}{1,2} and
\authorb{Milan S. Dimitrijevi{\' c}}{2,3,4}
\end{authorl}

\begin{addressl}
\addressb{1}{Institute of Physics, University of Belgrade,\\  P.O. Box 57,
11001 Belgrade, Serbia;}
\addressb{2}{Isaac Newton Institute of Chile, Yugoslavia Branch,\\ Volgina 7,
       11060 Belgrade Serbia;}
\addressb{3}{Astronomical Observatory, Volgina 7, 11060 Belgrade,
       Serbia;}
\addressb{4}{Observatoire de Paris, 92195 Meudon Cedex, France;}
\end{addressl}

\submitb{Received: 2011 December 2; accepted: 2008 December 15}

\begin{summary} In this work, the chemi-ionization processes
in atom- Rydberg atom collisions, as well as the corresponding
chemi-recombination processes are considered as factors of influence
on the atom exited-state populations in weakly ionized layers of
stellar atmospheres. The presented results are related to the
photospheres of the Sun and some M red dwarfs as well as weakly
ionized layers of DB white dwarfs atmospheres. It has been found
that the mentioned chemi ionization/recombination processes dominate
over the relevant concurrent electron-atom and electron-ion
ionization and recombination process in all parts of considered
stellar atmospheres. The obtained results demonstrate the fact that
the considered chemi ionization/recombination processes must have a
very significant influence on the optical properties of the stellar
atmospheres. Thus, it is shown that these processes and their
importance for non-local thermodynamic equilibrium (non-LTE) modeling of the
solar atmospheres should be investigated further.

\end{summary}


\resthead{Chemi-ionization/recombination processes in stelar
atmospheres} {Mihajlov et al.}

\sectionb{1}{INTRODUCTION}

Chemi-ionization processes in principle include the processes of
associative ionization
\begin{equation}
\label{eq:1a} A^{*}(n) + X  \Rightarrow   AX^{+} + e,
\end{equation}
as well as the processes of Penning ionization
\begin{equation}
\label{eq:1b} A^{*}(n) + X  \Rightarrow  A + X^{+} + e,
\end{equation}
where $A$, $X$ and $X^{+}$ are atoms and the atomic ions in their
ground states, A$^{*}$(n)- atom in a highly excited (Rydberg) state
with the principal quantum number $n \gg 1$, and $AX^{+}$ is the
corresponding molecular ion in its ground electronic state. It is
customary that in the case $A=X$ chemi-ionization processes are
treated as symmetric, and in the case $A \neq X$ - as non-symmetric.
Concerning Penning-ionization processes let us draw attention to the
fact that for all the processes which can be described by the scheme
(\ref{eq:1b}), only those are treated as chemi-ionization which go
on in a similar way as the associative-ionization processes
(\ref{eq:1a}). Here, all chemi-ionization processes are
treated on the basis of dipole resonant mechanism, which was
introduced in the considerations in Smirnov \& Mihajlov (1971) and
was described in details in Ignjatovic \& Mihajlov (2005) and
Ignjatovic et al. (2008).
The symmetric $A=X$ chemi-ionization processes
\begin{equation}
\label{eq:2a} A^{*}(n) + A  \Rightarrow   A_{2}^{+} + e,
\end{equation}
\begin{equation}
\label{eq:2b} A^{*}(n) + A  \Rightarrow A + A^{+} + e,
\end{equation}
and the corresponding inverse recombination processes,
\begin{equation}\label{eq:3a}
A_{2}^{+}+ e \Rightarrow A^{*}(n) + A,
\end{equation}
\begin{equation}\label{eq:3b}
A + A^{+}+ e \Rightarrow A^{*}(n) + A,
\end{equation}
where $A=H(1s)$ or $He(1s^2)$, were theoretically considered in
Refs. Mihajlov et al. (1997b) and Mihajlov et al. (1996) as
factors of influence on the populations of excited atoms in the
weakly ionized hydrogen and helium plasmas. It means that the
efficiency of these processes had to be compared with the efficiency
of the processes
\begin{equation}
\label{eq:3i} A^{*}(n) +  e \Rightarrow   A^{+} + 2 e,
\end{equation}
\begin{equation}
\label{eq:4r} A^{+} + 2 e  \Rightarrow  A^{*}(n) + e,
\end{equation}
\begin{equation}
\label{eq:5r} A^{+}+ e \Rightarrow A^{*}(n) + \varepsilon_{\lambda},
\end{equation}
where $A=H(1s)$ or $He(1s^2)$ and $\varepsilon_{\lambda}$ is the
energy of a photon with wavelength $\lambda$.  For the considered
conditions in Refs.~Mihajlov et al. (1997b) and Mihajlov et al.
(1996) the rate coefficients for all chemi-ionization and
chemi-recombination processes (\ref{eq:2a})-(\ref{eq:3b}) were
determined. It was found that under the mentioned conditions these
processes in the region $n \leq 10$ are dominant or at least
comparable with the concurrent electron-atom and electron-ion
processes (from the aspect of their influence on excited-atom
populations) when the ionization degree of the considered plasma is
$\leq 10^{-3}$.


It was just these results, as well as the experience gained earlier
with radiation ion-atom processes (Mihajlov \& Dimitrijevic (1986,
1992), etc.), that suggested the idea that the chemi-ionization
processes (\ref{eq:2a}), (\ref{eq:2b}) and the chemi-recombination
processes (\ref{eq:3a}), (\ref{eq:3b}) should be of considerable
interest \textbf{from the aspect of their influence on excited-atom
populations} for the weakly ionized layers of stellar atmospheres.
This was proven at a qualitative level for the hydrogen case (solar
photosphere) in Mihajlov et al. (1997a), and for the helium case
(atmospheres of DB white dwarfs with $T_{eff}= 12000 \div 18 000 K$)
in Mihajlov et al. (2003a).


Than, the influence of the chemi-ionization processes (\ref{eq:2a}),
(\ref{eq:2b}) and the chemi-recombination processes (\ref{eq:3a}),
(\ref{eq:3b}) with $A=H$ on the excited hydrogen atom populations
was examined in much more detail in Mihajlov et al. (2003b), where
these processes were included ab initio in a non-LTE modeling of an
M red dwarf atmosphere with the effective temperature $T_{eff} =
3800$ K, using PHOENIX code (see Baron \& Hauschildt (1998)). A fact
was established that including even the
chemi-ionization/recombination only for $4\le n \le 8$, generates
significant changes (by up to 50 percent), at least in the
populations of hydrogen-atom excited states with $2 \le n \le 20$.

Later, again using the PHOENIX code for the case of the atmosphere
of the same red dwarf, the influence was examined of the processes
(\ref{eq:2a}-\ref{eq:3b}) with $n \leq 10$ on the free electron
density and the profiles of hydrogen atom spectral lines. It was
established that if all these processes (with $n \ge 2$) are
included, a significant change (somewhere up to 2 - 3 times) for the
free electron density $N_{e}$, is also generated and, as one of
further consequences, significant changes in hydrogen line profiles.


In this paper one of our main aims is to draw attention to the
importance of all processes (\ref{eq:1a}) - (\ref{eq:2b}) with A = H
for non-LTE modeling of the solar atmosphere. For this purpose, it
should be demonstrated that in the solar photosphere the efficiency
of these processes is greater than, or at least comparable to, the
efficiency of processes (\ref{eq:3i}) - (\ref{eq:5r}) with A = H
within those ranges of values of $n \ge 2$ and temperature $T$ which
are relevant to the chosen solar atmosphere model. However, until
now only, for chemi-recombination processes (\ref{eq:2a}) and
(\ref{eq:2b}) was qualitatively shown that for $4\le n \le 8$ their
efficiency is comparable to the efficiency of the concurrent
processes (\ref{eq:4r}) and (\ref{eq:5r}) in a part of the solar
photosphere (see Mihajlov et al. (1997a)).

Besides all mentioned, the fact that the processes (\ref{eq:1a}) -
(\ref{eq:2b}) can be important for the solar photosphere is
supported by the results obtained in Mihajlov et al. (2003a, 2007)
in connection with an M red dwarf atmosphere ($T_{eff} = 3800$ K).

\sectionb{2}{THE RATE COEFFICIENTS OF THE
CHEMI-IONIZATION/RECOMBINATION PROCESSES}

The corresponding partial rate coefficients of the chemi-ionization
processes (\ref{eq:2a}) and (\ref{eq:2b}) are denoted here with
$K_{ci}^{(a,b)}(n;T)$, where $T$ is the temperature of the
considered plasma, and the partial rate coefficients of the inverse
chemi-recombination processes (\ref{eq:3a}) and (\ref{eq:3b}) - with
$K_{cr}^{(a,b)}(n;T)$. Under the conditions which exist in the solar
atmosphere, we can determine the rate coefficients
$K_{cr}^{(a,b)}(n;T)$ over the rate coefficients
$K_{ci}^{(a,b)}(n;T)$ from the principle of the thermodynamical
balance for processes (\ref{eq:2a}), (\ref{eq:2b}), (\ref{eq:3a})
and (\ref{eq:3b}) in the form

\begin{equation}\label{eq:Krb}
K_{ci}^{(a,b)}\cdot N_{n} N_{1} = K_{cr}^{(a,b)}(n,T)\cdot
N_{1}N_{ai}N_{e},
\end{equation}

\noindent where $N_{1}$ and $N_{n}$ denote the densities of ground- and
excited-state hydrogen atoms respectively. Using these partial rate
coefficients we will determine the total ones, namely,
\begin{equation}\label{eq:Kzbir}
K_{ci,cr}(n,T)=K_{ci,cr}^{(a)}(n,T)+K_{ci,cr}^{(b)}(n,T),
\end{equation}
which characterizes the efficiency of the considered
chemi-ionization and chemi-recombination processes together.

Here we will consider processes (\ref{eq:2a}) - (\ref{eq:3b}) with A
= H within the regions $n\ge 5$ and $2\le n \le 4$ separately. The
reason for it is the behavior of the adiabatic potential curves of
atom-atom systems $H^{*}(n)+H(1s)$. Namely, in the first region the
atom-atom curves lie above the adiabatic curve of the ion-ion system
$H^{+} + H^{-}(1s^{2})$ for any $R$, and the dipole resonant
mechanism can be applied for $n\ge 5$ without any exceptions.
Consequently, for $n\ge5$ we will determine the rate coefficients of
these processes following the previous papers (Mihajlov et al.
1997a, 2003a,b, 2007).

However, in the region $n\le4$ there are points where the atom-atom
curves cross the ion-ion one and application of this mechanism
generates some errors (see Janev \& Mihajlov (1979)).

Due to this fact and the mentioned errors, we use here
semi-empirical rate coefficients $K_{cr,ci}^{(a)}(n=3,T)$ and
$K_{cr,ci}^{(a)}(n=4,T)$, which are obtained on the bases of the
data from Janev et al. (1987). For relatively minor
chemi-ionization/re\-com\-bi\-nation processes with $n=2$ we use here rate
coefficients $K_{ci}^{(a)}(n=2,T)$ and $K_{cr}^{(a)}(n=2,T)$, which
are $10$ - $30$ percent greater than the corresponding coefficients
obtained using the data from Janev et al. (1987), in accordance with
the calculated results from Urbain et al. (1991) and Rawlings et al.
(1993). It gives a possibility to compensate the decrease of the
rate coefficients $K_{ci}^{(a)}(n\ge 5,T)$ and $K_{cr}^{(a)}(n\ge
5,T)$ in comparison with the corresponding ones obtained using Janev
et al. (1987), due to the fact that here, unlike Janev et al.
(1987), the decay of the considered system's initial electronic
state has been taken into account.

\sectionb{3}{RESULTS AND DISCUSSION}

In accordance with the aim of this work we consider here model C of
solar atmosphere from Vernazza et al. (1981). Namely, this is a
non-LTE model which is still actual (see Stix (2002)), and it is
only for this model that all the quantities necessary for our
calculations are available in tabular form as functions of height
($h$) in Solar photosphere.

Let $I_{ci}(n,T)$, $I_{cr}(n,T)$ be the total chemi-ionization and
chemi-recombination fluxes caused by the processes
(\ref{eq:2a},\ref{eq:2b}) and (\ref{eq:3a},\ref{eq:3b}), i.e.,
\begin{equation}\label{eq:I12}
I_{ci}(n,T) = K_{ci}(n,T)\cdot N_{n}N_{1}, \qquad I_{cr}(n,T) =
K_{cr}(n,T) \cdot N_{1}N_{i}N_{e},
\end{equation}
and $I_{i;ea}(n,T)$, $I_{r;eei}(n,T)$ and $I_{r;ph}(n,T)$ be the
fluxes caused by ionization and recombination processes
(\ref{eq:3i}), (\ref{eq:4r}) and (\ref{eq:5r}), i.e.
\begin{equation}\label{eq:I345}
I_{i;ea}(n,T) = K_{ea}(n,T)\cdot N_{n}N_{e}, \quad I_{r;eei}(n,T) =
K_{eei}(n,T) \cdot N_{i}N_{e}N_{e}, \quad I_{r;ph}(n,T) =
K_{ph}(n,T)\cdot N_{i}N_{e},
\end{equation}
where $N_{1}$, $N_{n}$, $N_{i}$, and $N_{e}$ are, respectively,
the densities of the ground and excited states of a hydrogen atom,
of ion $H^{+}$, and of free electron in the considered plasma with
given $T$.

\begin{figure}[ht]
\begin{minipage}[b]{0.5\linewidth}
\centering
\includegraphics[width=\columnwidth,
height=0.75\columnwidth]{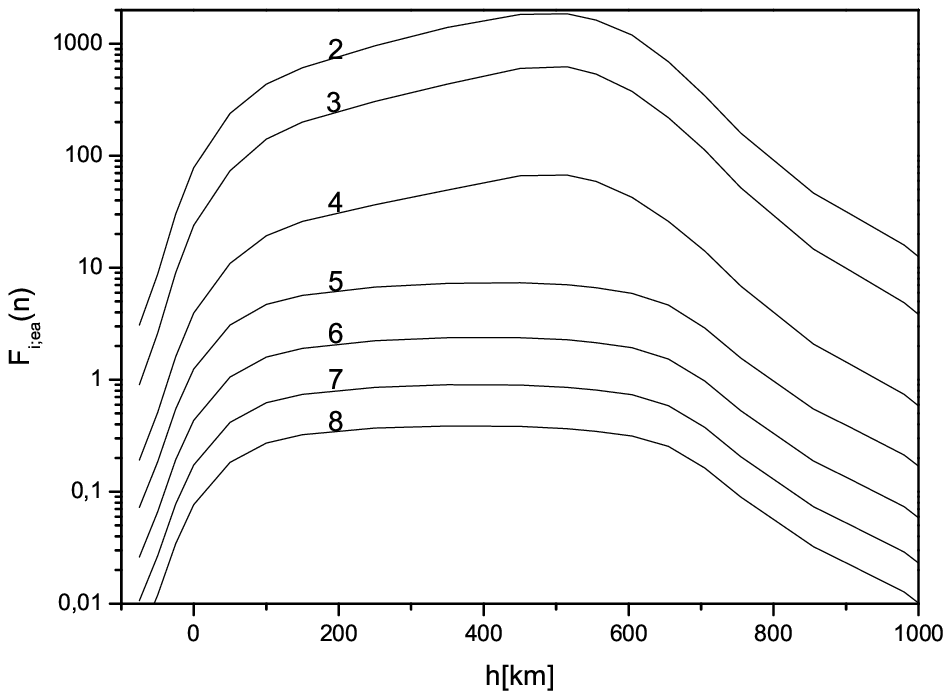} \caption{Behavior of the
quantity $F_{i;ea}^{(ab)}(n)$ given by Eq.~(\ref{eq:Fin}), as a
function of height h.} \label{fig:Fi_eaN}
\end{minipage}
\hspace{0.5cm}
\begin{minipage}[b]{0.5\linewidth}
\centering
\includegraphics[width=\columnwidth,
height=0.75\columnwidth]{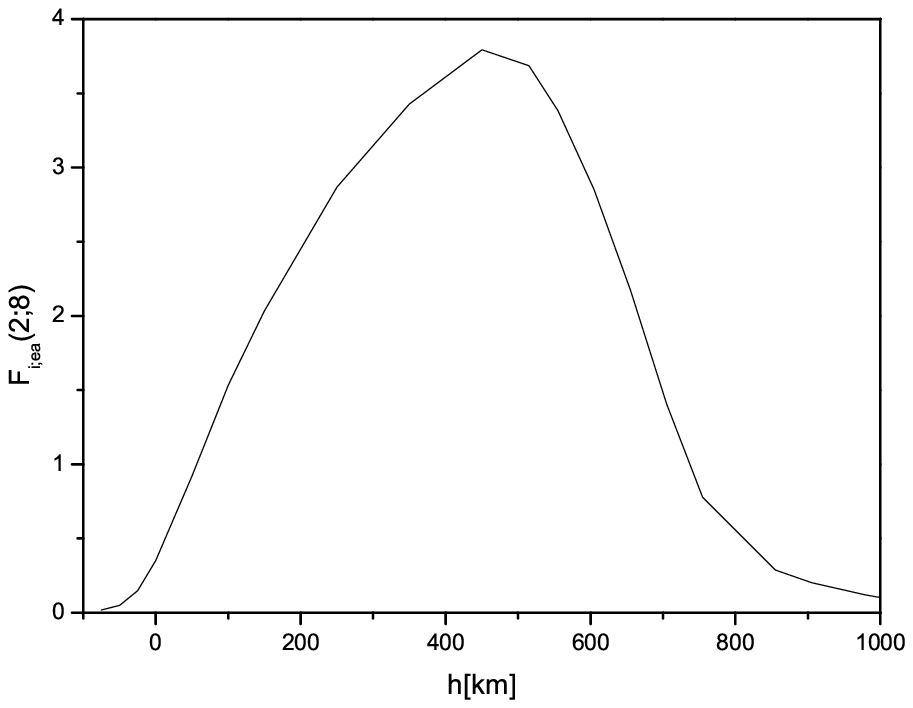} \caption{Behavior of the
quantity $F_{i;ea}(2;8)$ given by Eq.~(\ref{eq:Fiea2-8}), as a
function of height h.} \label{fig:Fi_ea}
\end{minipage}
\end{figure}

Using these expressions, we will first calculate quantities
$F_{i,ea}(n,T)$ given by
\begin{equation}\label{eq:Fin}
F_{i,ea}(n,T) = \frac{I_{ci}(n,T)}{I_{i;ea}(n,T)} = \frac
{K_{ci}(n,T)}{K_{ea}(n,T)}\cdot {N_{1}}{N_{e}},
\end{equation}
which characterize the relative efficiency of partial
chemi-ionization processes (\ref{eq:2a}) and (\ref{eq:2b}) together
and the impact electron-atom ionization (\ref{eq:3i}) in the
considered plasma. The impact ionization rate coefficients
$K_{ea}(n,T)$ are taken from Vriens \& Smeets(1980). In Figure
\ref{fig:Fi_eaN} the behavior of the quantities $F_{i,ea}(n,T)$ for
$2\le n \le 8$ as functions of height $h$ is shown. One can see that
the efficiency of the considered chemi-ionization processes in
comparison with the electron-atom impact ionization is dominant for
2$\le n \le$6 and becomes comparable for $n = 7$ and $8$.

Than, in order to compare the relative influence of the
chemi-ionization processes (\ref{eq:2a}) and (\ref{eq:2b}) together
to that of the impact electron-atom ionization process (\ref{eq:3i})
on the whole block of the excited hydrogen atom states with $2\le n
\le 8$, we will calculate  quantity $F_{i,ea;2-8}(T)$, given by
\begin{equation}\label{eq:Fiea2-8}
F_{i,ea;2-8}(T) = \frac{\sum\limits_{n=2}^{8}
I_{ci}(n,T)}{\sum\limits_{n=2}^{8} I_{i;ea}(n,T)} = \frac
{\sum\limits_{n=2}^{8} K_{ci}(n,T)\cdot N_{n}}{\sum\limits_{n=2}^{8}
K_{ea}(n,T)\cdot N_{n}}\cdot {N_{1}}{N_{e}},
\end{equation}
which can reflect the influence of the existing populations of
excited hydrogen atom states within a non-LTE model of solar
atmosphere. In Figure \ref{fig:Fi_ea} the behavior of the quantity
$F_{i,ea;2-8}(T)$ as functions of height $h$ is shown. As one can
see, the real influence of the chemi-ionization processes on the
total populations of states with $2 \le n \le 8$ remains dominant
with respect to the concurrent electron-atom impact ionization
processes almost in the whole photosphere (50 km $\leq h \leq$ 750
km).

Finally, we compared the relative influence of the
chemi-recombination processes (\ref{eq:3a}) and (\ref{eq:3b})
together and total influence of the electron - electron - $H^{+}$
recombination process (\ref{eq:4r}) and photo-recombination electron
- $H^{+}$ process (\ref{eq:5r}) on the same block of excited
hydrogen atom states with $2\le n \le 8$. It was confirmed a
domination of the chemi-recombination processes with $2 \le n \le 8$
over the mentioned concurrent processes in a significant part of the
photosphere (-50 km $\leq h \leq$ 600 km).

%
%

The obtained results demonstrate the fact that the considered
chemi-ionization/re\-com\-bi\-on\-ti\-on processes must have a very
significant influence on the optical properties of the solar
photosphere in comparison to the concurrent electron-atom impact
ionization and electron-ion recombination processes. Thus it is
shown that the importance of these processes for non-LTE modeling of
solar atmosphere should be necessarily investigated.

\thanks{ This work was supported by the Ministry of Science and Technological
Development of Serbia as a part of the projects 176002 and III4402.}

\References

\refb Baron, E. \& Hauschildt, P.~H. 1998, ApJ, 495, 370

\refb Ignjatovi\'{c}, L.~M. \& Mihajlov, A.~A. 2005, Phys. Rev. A.,
72, 022715

\refb Ignjatovi{\'c}, L.~M., Mihajlov, A.~A., \& Klyucharev, A.~N.\
2008, J.Phys.B Atomic Molecular Physics, 41, 025203

\refb Janev, R.~K., Langer, W.~D., Evans Jr., K., \& Post Jr., D. E.
1987, { \it Elementary Processes in Hydrogen-Helium Plasmas}
(Springer-Verlag)

\refb Janev, R.~K. \& Mihajlov, A.~A. 1979, Phys. Rev. A., 20, 1890

\refb Mihajlov, A.~A. \& Dimitrijevi\'c, M.~S. 1986, A\&A, 155,
  319

\refb Mihajlov, A.~A. \& Dimitrijevi\'c, M.~S. 1992, A\&A, 256, 305

\refb Mihajlov, A.~A., Dimitrijevi\'c, M.~S., \& Djuri\'c, Z. 1996,
Physica Scripta, 53, 159

\refb Mihajlov, A.~A., Djuric, Z., Dimitrijevic, M.~S., \&
Ljepojevic, N.~N.\ 1997b, Physica Scripta, 56, 631

\refb Mihajlov, A.~A., Ignjatovic, L.~M., Vasilijevi\'c, M.~M., \&
Dimitrijevi\'c, M.~S. 1997a, A\&A, 324, 1206

\refb Mihajlov, A.~A., Ignjatovi\'c, L.~M., Dimitrijevi\'c, M.~S.,
\& Djuri\'c, Z. 2003a, ApJS, 147, 369

\refb Mihajlov, A.~A., Jevremovi\'c, D., Hauschildt, P., Dimitrijevi\' c M.S.,
Ignjatovi\'c, \& Alard 2003b, A\&A, 403, 787

\refb Mihajlov, A.~A., Jevremovi\'c, D., Hauschildt, P., Dimitrijevi\' c M.S.,
Ignjatovi\'c, \& Alard 2007, A\&A, 471, 671

\refb Mihajlov, A.~A., Ljepojevic, N.~N., \& Dimitrijevic, M.~S.\
1992, J.Phys.B Atomic Molecular Physics, 25, 5121

\refb Rawlings, J., Drew, J., \& Barlow, M. 1993, MNRAS, 265, 968

\refb Smirnov, V.~M. \& Mihajlov, A.~A. 1971, Opt. Spektrosk., 30,
984

\refb Stix, M. 2002, {\it The Sun} (Section 4.3 Atmospheric Models)
(Springer Verlag, 2nd)

\refb Urbain, X., Cornet, A., Brouillard, F., \& Giusti-Suzor, A.
1991, Phys.Rev.Lett., 66, 1685

\refb Vernazza, J., Avrett, E., \& Loser, R. 1981, ApJS, 45, 635

\refb Vriens, L. \& Smeets, A. H.~M. 1980, Phys. Rev. A, 22, 940

\end{document}